# Neutrosophic Relational Data Model

Haibin wang

Biostatistics Research and Informatics Core

Winship Cancer Institute

Emory University

Atlanta, GA, 30322

haibin.wang@emoryhealthcare.org

Rajshekhar Sunderraman

Department of Computer Science

Georgia State University

Atlanta, GA, 30303

raj@cs.gsu.edu

Florentin Smarandache

Department of Mathematics

University of New Mexico

Gallup, NM 87301

smarand@unm.edu

André Rogatko

Biostatistics Research and Informatics Core

Winship Cancer Institute

Emory University

Atlanta, GA, 30322

andre.rogatko@emoryhealthcare.org

## Abstract

In this paper, we present a generalization of the relational data model based on interval neutrosophic set [1]. Our data model is capable of manipulating incomplete as well as inconsistent information. Fuzzy relation or intuitionistic fuzzy relation can only handle incomplete information. Associated with each relation are two membership functions one is called truth-membership function $T$ which keeps track of the extent to which we believe the tuple is in the relation, another is called falsity-membership function $F$ which keeps track of the extent to which we believe that it is not in the relation. A neutrosophic relation is inconsistent if there exists one tuple α such that *T*(α) + *F*(α) > 1 . In order to handle inconsistent situation, we propose an operator called "split" to transform inconsistent neutrosophic relations into pseudo-consistent neutrosophic relations and do the set-theoretic and relation-theoretic operations on them and finally use another operator called "combine" to transform the result back to neutrosophic relation. For this data model, we define algebraic operators that are generalizations of the usual operators such as intersection, union, selection, join on fuzzy relations. Our data model can underlie any database and knowledge-base management system that deals with incomplete and inconsistent information.

**Keyword**: Interval neutrosophic set, fuzzy relation, inconsistent information, incomplete information, neutrosophic relation.

1. Introduction

Relational data model was proposed by Ted Codd's pioneering paper [2]. Since then, relational database systems have been extensively studied and a lot of commercial relational database systems are currently available [3, 4]. This data model usually takes care of only well-defined and unambiguous data. However, imperfect information is ubiquitous – almost all the information that we have about the real world is not certain, complete and precise [5]. Imperfect information can be classified as: incompleteness, imprecision, uncertainty, and inconsistency. Incompleteness arises from the absence of a value, imprecision from the existence of a value which cannot be measured with suitable precision, uncertainty from the fact that a person has given a subjective opinion about the truth of a fact which he/she does not know for certain, and inconsistency from the fact that there are two or more conflicting values for a variable.

In order to represent and manipulate various forms of incomplete information in relational databases, several extensions of the classical relational model have been proposed [6, 7, 8, 9, 10, 11]. In some of these extensions, a variety of "null values" have been introduced to model unknown or not-applicable data values. Attempts have also been made to generalize operators of relational algebra to manipulate such extended data models [6, 8, 11, 12, 13]. The fuzzy set theory and fuzzy logic proposed by Zadeh [14] provide a requisite mathematical framework for dealing with incomplete and imprecise information. Later on, the concept of interval-valued fuzzy sets was proposed to capture the fuzziness of grade of membership itself [15]. In 1986, Atanassov introduced the intuitionistic fuzzy set [16] which is a generalization of fuzzy set and provably equivalent to interval-valued fuzzy set. The intuitionistic fuzzy sets consider both truth-membership *T* and falsity-membership *F* with $T(a), F(a) \in [0,1]$ and $T(a) + F(a) \leq 1$. Because of the restriction, the fuzzy set, interval-valued fuzzy set, and intuitionistic fuzzy set cannot handle inconsistent information. Some authors [17, 18, 19, 20, 21, 22, 23] have studied relational databases in the light of fuzzy set theory with an objective to accommodate a wider range of real-world requirements and to provide closer man-machine interactions. Probability, possibility, and Dempster-Shafer theory have been proposed to deal with uncertainty. Possibility theory [24] is built upon the idea of a fuzzy restriction. That means a variable could only take its value from some fuzzy set of values and any value within that set is a possible value for the variable. Because values have different degrees of membership in the set, they are possible to different degrees. Prade and Testemale [25] initially suggested using possibility theory to deal with incomplete and uncertain information in database. Their work is extended in [26] to cover multivalued attributes. Wong [27] proposes a method that quantifies the uncertainty in a database using probabilities. His method maybe is the simplest one which attached a probability to every member of a relation, and to use these values to provide the probability that a particular value is the correct answer to a particular query. Carvallo and Pittarelli [28] also use probability theory to model uncertainty in relational databases systems. Their method augmented projection and join operations with probability measures.

However, unlike incomplete, imprecise, and uncertain information, inconsistent information has not enjoyed enough research attention. In fact, inconsistent information exists in a lot of applications. For example, in data warehousing application, inconsistency will appear when trying to integrate the data from many different sources. Another example is that in the expert system, there exist facts which are inconsistent with each other. Generally, two basic approaches have been followed in solving the inconsistency problem in knowledge base: belief revision and paraconsistent logic. The goal of the first approach is to make an inconsistent theory consistent, either by revising it or by representing it by a consistent semantics. On the other hand, the paraconsistent approach allows reasoning in the presence of inconsistency, and contradictory information can be derived or introduced without trivialization [29]. Bagai and Sunderraman [30, 31] proposed a paraconsistent realational data model to deal with incomplete and inconsistent information. The data model has been applied to compute the well-founded and fitting model of logic programming [32, 33]. This data model is based on paraconsistent logics which were studied in detail by de Costa [34] and Belnap [35].

In this paper, we present a new relational data model – neutrosophic relational data model (NRDM). Our model is based on the neutrosophic set theory which is an extension of intuitionistic fuzzy set theory [36] and is capable of manipulating incomplete as well as inconsistent information. We use both truth-membership function grade α and falsity-membership function grade β to denote the status of a tuple of a certain relation with $\alpha, \beta \in [0,1]$ and $\alpha + \beta \leq 2$. NRDM is the generalization of fuzzy relational data model (FRDM). That is , when α + β = 1, neutroshophic relation is the ordinary fuzzy relation. This model is distinct with paraconsistent relational data model (PRDM), in fact it can be easily shown that PRDM is a special case of NRDM. That is, when α, β = 0 or 1, neutrosophic relation is just paraconsistent relation. We can use Figure 1 to express the relationship among FRDM, PRDM, and NRDM.

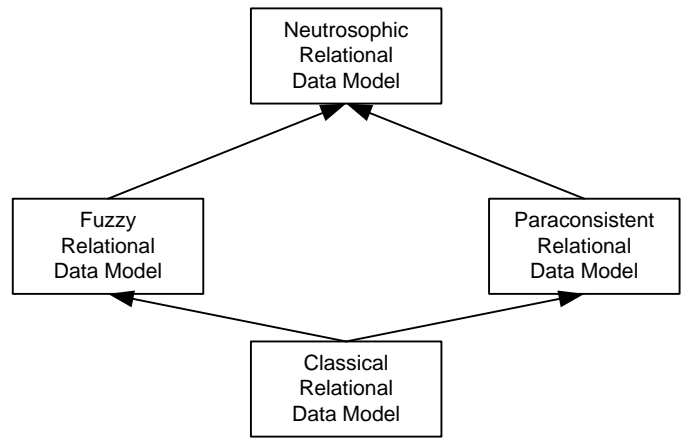

Figure 1. Relationship among RDM, FRDM, PRDM, and NRDM

We introduce neutrosophic relations, which are the fundamental mathematical structures underlying our model. These structures are strictly more general than classical fuzzy relations and intuititionistic fuzzy relations (interval-valued fuzzy relations), in that for any fuzzy relation or intuitionistic fuzzy relation there is a neutrosophic relation with the same information content, but not *vice versa*. The claim is also true for the relationship between neutrosophic relations and paraconsistent relations. We define algebraic operators over neutrosophic relations that extend the standard operators such as selection, join, union over fuzzy relations.

There are many potential applications of our new data model. Here are some examples:

a) Web mining. Essentially the data and documents on the Web are heterogeneous, inconsistency is unavoidable. Using the presentation and reasoning method of our data model, it is easier to capture imperfect information on the Web which will provide more potentially valued-added information.

b) Bioinformatics. There is a proliferation of data sources. Each research group and each new experimental technique seems to generate yet another source of valuable data. But these data can be incomplete and imprecise, and even inconsistent. We could not simply throw away one data in favor of other data. So how to represent and extract useful information from these data will be a challenge problem.

c) Decision Support System. In decision support system, we need to combine the database with the knowledge base. There will be a lot of uncertain and inconsistent information, so we need an efficient data model to capture these information and reasoning with these information.

The paper is organized as follow. Section 2 deals with some of the basic definitions and concepts of fuzzy relations and operations. Section 3 introduces neutrosophic relations and two notions of generalizing the fuzzy relational operators such as union, join, projection for these relations. Section 4 presents some actual generalized algebraic operators for the neutrosophic relations. These operators can be used for specifying queries for database systems built on such relations. Section 5 gives an illustrative application of these operators. Finally, section 6 contains some concluding remarks and directions for future work.

## 2. Fuzzy Relations and Operations

In this section, we present the essential concepts of a fuzzy relational database. Fuzzy relations associate a value between 0 and 1 with every tuple representing the degree of membership of the tuple in the relation. We also present several useful query operators on fuzzy relations.

Let a *relation scheme* (or just *scheme*) $\Sigma$ be a finite set of *attribute names*, where for any attribute name $A \in \Sigma$, dom(A) is a non-empty *domain* of values for A. *A tuple* on $\Sigma$ is any map $t : \Sigma \to \bigcup_{A \in \Sigma} dom(A)$, such that $t(A) \in dom(A)$, for each $A \in \Sigma$. Let $\tau(\Sigma)$ denote the set of all tuples on $\Sigma$.

**Definition 1**  A *fuzzy relation* on scheme $\Sigma$ is any map $R : \tau(\Sigma) \to [0,1]$. We let $F(\Sigma)$ be the set of all fuzzy relations on $\Sigma$.

If $\Sigma$ and $\Delta$ are relation schemes such that $\Delta \subseteq \Sigma$, then for any tuple $t \in \tau(\Delta)$, we let $t^\Sigma$ denote the set $\{t' \in \tau(\Sigma) \mid t'(A) = t(A), \text{for all } A \in \Delta\}$ of all extensions of $t$. We extend this notion for any $T \subseteq \tau(\Delta)$ by defining $T^\Sigma = \bigcup_{t \in T} t^\Sigma$.

### 2.1 Set-theoretic operations on Fuzzy relations

**Definition 2 Union:** Let $R$ and $S$ be fuzzy relations on scheme $\Sigma$. Then, $R \cup S$ is a fuzzy relation on scheme $\Sigma$ given by

$$(R \cup S)(t) = \max\{R(t), S(t)\}, \text{for any } t \in \tau(\Sigma).$$

**Definition 3 Complement**: Let $R$ be a fuzzy relation on scheme $\Sigma$. Then, $-R$ is a fuzzy relation on scheme $\Sigma$ given by

$$(-R)(t) = 1 - R(t), \text{ for any } t \in \tau(\Sigma).$$

**Definition 4 Intersection**: Let $R$ and $S$ be fuzzy relations on scheme $\Sigma$. Then $R \cap S$ is a fuzzy relation on scheme $\Sigma$ given by

$$(R \cap S)(t) = \min\{R(t), S(t)\}, \text{ for any } t \in \tau(\Sigma).$$

**Definition 5 Difference**: Let $R$ and $S$ be fuzzy relations on scheme $\Sigma$. Then, $R - S$ is a fuzzy relation on scheme $\Sigma$ given by

$$(R - S)(t) = \min\{R(t), 1 - S(t)\}, \text{ for any } t \in \tau(\Sigma).$$

## 2.2 Relation-theoretic operations on Fuzzy relations

**Definition 6** Let $R$ and $S$ be fuzzy relations on schemes $\Sigma$ and $\Delta$, respectively. Then, the *natural join* (or just *join*) of $R$ and $S$, denoted $R \infty S$ is a fuzzy relation on scheme $\Sigma \cup \Delta$, given by

$$(R \infty S)(t) = \min\{R(\pi_\Sigma(t)), S(\pi_\Delta(t))\}, \text{ for any } t \in \tau(\Sigma \cup \Delta).$$

**Definition 7** Let $R$ be a fuzzy relation on scheme $\Sigma$ and let $\Delta \subseteq \Sigma$. Then, the *projection* of $R$ onto $\Delta$, denoted by $\prod_\Delta(R)$ is a fuzzy relation on scheme $\Delta$ given by

$$(\prod_\Delta(R))(t) = \max\{R(u) \mid u \in t^\Sigma\}, \text{ for any } t \in \tau(\Delta).$$

**Definition 8** Let $R$ be a fuzzy relation on scheme $\Sigma$, and let $F$ be any logic formula involving attribute names in $\Sigma$, constant symbols (denoting values in the attribute domains), equality symbol $=$, negation symbol $\neg$, and connectives $\vee$ and $\wedge$. Then, the selection of $R$ by $F$, denoted $\overset{\bullet}{\sigma}_F(R)$, is a fuzzy relation on scheme $\Sigma$, given by

$$(\overset{\bullet}{\sigma}_F(R))(t) = \begin{cases} R(t) & \text{if } t \in \sigma_F(\tau(\Sigma)) \\ 0 & \text{otherwise} \end{cases}$$

where $\sigma_F$ is the usual selection of tuples satisfying $F$.

## 3. Neutrosophic Relations

In this section, we generalize fuzzy relations in such a manner that we are now able to assign a measure of belief and a measure of doubt to each tuple. We shall refer to these generalized fuzzy relations as *neutrosophic relations*. So, a tuple in a neutrosophic relation is assigned a measure $\langle \alpha, \beta \rangle, 0 \leq \alpha, \beta \leq 1$. $\alpha$ will be referred to as the *belief* factor and $\beta$ will be referred to as the *doubt* factor. The interpretation of this measure is that we believe with confidence $\alpha$ and doubt with confidence $\beta$ that the tuple is in the relation. The belief and doubt confidence factors for a tuple need not add to exactly 1. This allows for incompleteness and inconsistency to be represented. If the belief and doubt factors add up to less than 1, we have incomplete information regarding the tuple's status in the relation and if the belief and doubt factors add up to more than 1, we have inconsistent information regarding the tuple's status in the relation.

In contrast to fuzzy relations where the grade of membership of a tuple is fixed, neutrosophic relations bound the grade of membership of a tuple to a subinterval $[\alpha, 1-\beta]$ for the case $\alpha + \beta \leq 1$.

The operators on fuzzy relations can also be generalized for neutrosophic relations. However, any such generalization of operators should maintain the belief system intuition behind neutrosophic relations.

This section also develops two different notions of operator generalizations.

We now formalize the notion of a neutrosophic relation.

Recall that $\tau(\Sigma)$ denotes the set of all tuples on any scheme $\Sigma$.

**Definition 9** A *neutrosophic relation* $R$ on scheme $\Sigma$ is any subset of

$$\tau(\Sigma) \times [0,1] \times [0,1]$$

For any $t \in \tau(\Sigma)$, we shall denote an element of $R$ as $\langle t, R(t)^+, R(t)^- \rangle$, where $R(t)^+$ is the belief factor assigned to $t$ by $R$ and $R(t)^-$ is the doubt factor assigned to $t$ by $R$. Let $V(\Sigma)$ be the set of all neutrosophic relations on $\Sigma$.

**Definition 10** A neutrosophic relation $R$ on scheme $\Sigma$ is *consistent* if $R(t)^+ + R(t)^- \leq 1$, for all $t \in \tau(\Sigma)$. Let $C(\Sigma)$ be the set of all consistent neutrosophic relations on $\Sigma$. $R$ is said to be *complete* if $R(t)^+ + R(t)^- \geq 1$, for all $t \in \tau(\Sigma)$. If $R$ is both consistent and complete, *i.e.*

$R(t)^+ + R(t)^- = 1$, for all $t \in \tau(\Sigma)$, then it is a *total* neutrosophic relation, and let $T(\Sigma)$ be the set of all total neutrosophic relations on $\Sigma$.

**Definition 11** $R$ is said to be *pseudo-consistent* if $\max\{b_i \mid (\exists t \in \tau(\Sigma))(\exists d_i)(\langle t, b_i, d_i \rangle \in R)\} + \max\{d_i \mid (\exists t \in \tau(\Sigma))(\exists b_i)(\langle t, b_i, d_i \rangle \in R\} > 1$, where for these $\langle t, b_i, d_i \rangle, b_i + d_i = 1$. Let $P(\Sigma)$ be the set of all pseudo-consistent neutrosophic relations on $\Sigma$.

**Example 1** Neutrosophic relation $R = \{\langle a, 0.3, 0.7 \rangle, \langle a, 0.4, 0.6 \rangle, \langle b, 0.2, 0.5 \rangle, \langle c, 0.4, 0.3 \rangle\}$ is pseudo-consistent. Because for $t = a, \max\{0.3, 0.4\} + \max\{0.7, 0.6\} = 1.1 > 1$.

It should be observed that total neutrosophic relations are essentially fuzzy relations where the uncertainty in the grade of membership is eliminated. We make this relationship explicit by defining a one-one correspondence $\lambda_\Sigma : T(\Sigma) \to F(\Sigma)$, given by $\lambda_\Sigma(R)(t) = R(t)^+$, for all $t \in \tau(\Sigma)$. This correspondence is used frequently in the following discussion.

### 3.1 Operator Generalizations

It is easily seen that neutrosophic relations are a generalization of fuzzy relations, in that for each fuzzy relation there is a neutrosophic relation with the same information content, but not *vice versa*. It is thus natural to think of generalizing the operations on fuzzy relations such as union, join, and projection etc. to neutrosophic relations. However, any such generalization should be intuitive with respect to the belief system model of neutrosophic relations. We now construct a framework for operators on both kinds of relations and introduce two different notions of the generalization relationship among their operators.

An $n$-ary operator on fuzzy relations with signature $\langle \Sigma_1, ..., \Sigma_{n+1} \rangle$ is a function $\Theta : F(\Sigma_1) \times \cdots \times F(\Sigma_n) \to F(\Sigma_{n+1})$, where $\Sigma_1, ... \Sigma_{n+1}$ are any schemes. Similarly, an $n$-ary operator on neutrosophic relations with signature $\langle \Sigma_1, ..., \Sigma_{n+1} \rangle$ is a function $\Psi : V(\Sigma_1) \times \cdots V(\Sigma_n) \to V(\Sigma_{n+1})$.

**Definition 12** An operator $\Psi$ on neutrosophic relations with signature $\langle \Sigma_1, ..., \Sigma_{n+1} \rangle$ is *totality preserving* if for any total neutrosophic relations $R_1, ..., R_n$ on schemes $\Sigma_1, ..., \Sigma_n$, respectively, $\Psi(R_1, ..., R_n)$ is also total.

**Definition 13** A totality preserving operator $\Psi$ on neutrosophic relations with signature $\langle \Sigma_1, ..., \Sigma_{n+1} \rangle$ is a *weak generalization* of an operator $\Theta$ on fuzzy relations with the same signature, if for any total neutrosophic relations $R_1,...,R_n$ on scheme $\Sigma_1,...,\Sigma_n$, respectively, we have

$$\lambda_{\Sigma_{n+1}}(\Psi(R_1,...,R_n)) = \Theta(\lambda_{\Sigma_1}(R_1),...,\lambda_{\Sigma_n}(R_n)).$$

The above definition essentially requires $\Psi$ to coincide with $\Theta$ on total neutrosophic realtions (which are in one-one correspondence with the fuzzy relations). In general, there may be many operators on neutrosophic relations that are weak generalizations of a given operator $\Theta$ on fuzzy relations. The behavior of the weak generalizations of $\Theta$ on even just the consistent neutrosophic relations may in general vary. We require a stronger notion of operator generalization under which, at least when restricted to consistent neutrosophic relations, the behavior of all the generalized operators is the same. Before we can develop such a notion, we need that of 'representation' of a neutrosophic relation.

We associate with a consistent neutrosophic relation $R$ the set of all (fuzzy relations corresponding to) total neutrosophic relations obtainable from $R$ by filling the gaps between the belief and doubt factors for each tuple. Let the map $reps_\Sigma : C(\Sigma) \to 2^{F(\Sigma)}$ be given by

$$reps_\Sigma(R) = \{Q \in F(\Sigma) \mid \bigwedge_{t_i \in \tau(\Sigma)} (R(t_i)^+ \leq Q(t_i) \leq 1 - R(t_i)^-)\}.$$

The set $reps_\Sigma(R)$ contains all fuzzy relations that are 'completions' of the consistent neutrosophic relation $R$. Observe that $reps_\Sigma$ is defined only for consistent neutrosophic relations and produces sets of fuzzy relations. Then we have following observation.

**Proposition 1** For any consistent neutrosophic relation $R$ on scheme $\Sigma$, $reps_\Sigma(R)$ is the singleton $\{\lambda_\Sigma(R)\}$ iff $R$ is total.

Proof  It is clear from the definition of consistent and total neutrosophic relations and from the definition of $reps$ operation.

We now need to extend operators on fuzzy relations to sets of fuzzy relations. For any operator $\Theta : F(\Sigma_1) \cdots F(\Sigma_n) \to F(\Sigma_{n+1})$ on fuzzy relations, we let $S(\Theta) : 2^{F(\Sigma_1)} \times \cdots \times 2^{F(\Sigma_n)} \to 2^{F(\Sigma_{n+1})}$ be a map on sets of fuzzy relations defined as follows. For any sets $M_1,...,M_n$ of fuzzy relations on schemes $\Sigma_1,...,\Sigma_n$, respectively,

$$S(\Theta)(M_1,...,M_n) = \{\Theta(R_1,...,R_n) \mid R_i \in M_i, \text{ for all } i, 1 \leq i \leq n\}.$$

In other words, $S(\Theta)(M_1,...,M_n)$ is the set of $\Theta$-images of all tuples in the Cartesian product $M_1 \times \cdots \times M_n$. We are now ready to lead up to a stronger notion of operator generalization.

**Definition 14**  An operator $\Psi$ on neutrosophic relations with signature $\langle \Sigma_1,...,\Sigma_{n+1} \rangle$ is *consistency preserving* if for any consistent neutrosophic relations $R_1,...,R_n$ on schemes $\Sigma_1,...,\Sigma_n$, respectively, $\Psi(R_1,...,R_n)$ is also consistent.

**Definition 15**  A consistency preserving operator $\Psi$ on neutrosophic relations with signature $\langle \Sigma_1,...,\Sigma_{n+1} \rangle$ is a *strong generalization* of an operator $\Theta$ on fuzzy relations with the same signature, if for any consistent neutrosophic relations $R_1,...,R_n$ on schemes $\Sigma_1,...,\Sigma_n$, respectively, we have

$$reps_{\Sigma_{n+1}}(\Psi(R_1,...,R_n)) = S(\Theta)(reps_{\Sigma_1}(R_1),...,reps_{\Sigma_n}(R_n)).$$

Given an operator $\Theta$ on fuzzy relations, the behavior of a weak generalization of $\Theta$ is 'controlled' only over the total neutrosophic relations. On the other hand, the behavior of a strong generalization is 'controlled' over all consistent neutrosophic relations. This itself suggests that strong generalization is a stronger notion than weak generalization. The following proposition makes this precise.

**Proposition 2**  If $\Psi$ is a strong generalization of $\Theta$, then $\Psi$ is also a weak generalization of $\Theta$.

Proof   Let $\langle \Sigma_1,...,\Sigma_{n+1} \rangle$ be the signature of $\Psi$ and $\Theta$, and let $R_1,...,R_n$ be any total neutrosophic relations on schemes $\Sigma_1,...,\Sigma_n$, respectively. Since all total relations are consistent, and $\Psi$ is a strong generalization of $\Theta$, we have that

$$reps_{\Sigma_{n+1}}(\Psi(R_1,...,R_n)) = S(\Theta)(reps_{\Sigma_1}(R_1),...,reps_{\Sigma_n}(R_n)),$$

Proposition 1 gives us that for each $i, 1 \leq i \leq n$, $reps_{\Sigma_i}(R_i)$ is the singleton set $\{\lambda_{\Sigma_i}(R_i)\}$. Therefore, $S(\Theta)(reps_{\Sigma_1}(R_i),...,reps_{\Sigma_n}(R_n))$ is just the singleton set: $\{\Theta(\lambda_{\Sigma_1}(R_1),...,\lambda_{\Sigma_n}(R_n))\}$. Here, $\Psi(R_1,...,R_n)$ is total, and $\lambda_{\Sigma_{n+1}}(\Psi(R_1,...,R_n)) = \Theta(\lambda_{\Sigma_1}(R_1),...,\lambda_{\Sigma_n}(R_n))$, i.e. $\Psi$ is a weak generalization of $\Theta$.

Though there may be many strong generalizations of an operator on fuzzy relations, they all behave the same when restricted to consistent neutrosophic relations. In the next section, we propose strong generalizations for the usual operators on fuzzy relations. The proposed generalized operators on neutrosophic relations correspond to the belief system intuition behind neutrosophic relations.

First we will introduce two special operators on neutrosophic relations called split and combine to transform inconsistent neutrosophic relations into pseudo-consistent neutrosophic relations and transform pseudo-consistent neutrosophic relations into inconsistent neutrosophic relations.

**Definition 16 (Split Operator $\Delta$)** Let $R$ be a neutrosophic relation on scheme $\Sigma$. Then,

$\Delta(R) = \{\langle t,b,d \rangle \mid \langle t,b,d \rangle \in R \text{ and } b+d \leq 1\} \cup$

$\{\langle t,b',d' \rangle \mid \langle t,b,d \rangle \in R \text{ and } b+d > 1 \text{ and } b' = b \text{ and } d' = 1-b\} \cup$

$\{\langle t,b',d' \rangle \mid \langle t,b,d \rangle \in R \text{ and } b+d > 1 \text{ } b' = 1-d \text{ and } d' = d\}.$

It is obvious that $\Delta(R)$ is pseudo-consistent if $R$ is inconsistent.

**Definition 17 (Combine Operator $\nabla$)** Let $R$ be a neutrosophic relation on scheme $\Sigma$. Then,

$\nabla(R) = \{\langle t,b',d' \rangle \mid (\exists b)(\exists d)((\langle t,b',d \rangle \in R \text{ and } (\forall b_i,d_i)(\langle t,b_i,d_i \rangle \rightarrow b' \geq b_i) \text{ and }$

$\langle t,b,d' \rangle \in R \text{ and } (\forall b_i)(\forall d_i)(\langle t,b_i,d_i \rangle \rightarrow d' \geq d_i))\}.$

It is obvious that $\nabla(R)$ is inconsistent if $R$ is pseudo-consistent.

Note that strong generalization defined above only holds for consistent or pseudo-consistent neutrosophic relations. For any arbitrary neutrosophic relations, we should first use split operation to transform them into non-inconsistent neutrosophic relations and apply the set-theoretic and relation-theoretic operations on them and finally use combine operation to transform the result into arbitrary neutrosophic relation. For the simplification of notation, the following generalized algebra is defined under such assumption.

## 4. Generalized Algebra on Neutrosophic Relations

In this section, we present one strong generalization each for the fuzzy relation operators such as union, join, and projection. To reflect generalization, a hat is placed over a fuzzy relation operator to obtain the corresponding neutrosophic relation operator. For example, $\infty$ denotes the natural join mong fuzzy relations, and $\hat{\infty}$ denotes natural join on neutrosophic relations. These generalized operators maintain the belief system intuition behind neutrosophic relations.

## 4.1 Set-Theoretic Operators

We first generalize the two fundamental set-theoretic operators, union and complement.

**Definition 18** Let $R$ and $S$ be neutrosophic relations on scheme $\Sigma$. Then,

(a) the *union* of $R$ and $S$, denoted $R \hat{\cup} S$, is a neutrosophic relation on scheme $\Sigma$, given by

$$(R \hat{\cup} S)(t) = \langle \max\{R(t)^+, S(t)^+\}, \min\{R(t)^-, S(t)^-\} \rangle, \text{ for any } t \in \tau(\Sigma);$$

(b) the *complement* of $R$, denoted $\hat{-}R$, is a neutrosophic relation on scheme $\Sigma$, given by

$$(\hat{-}R)(t) = \langle R(t)^-, R(t)^+ \rangle, \text{ for any } t \in \tau(\Sigma).$$

An intuitive appreciation of the union operator can be obtained as follows: Given a tuple $t$, since we believed that it is present in the relation $R$ with confidence $R(t)^+$ and that it is present in the relation $S$ with confidence $S(t)^+$, we can now believe that the tuple $t$ is present in the "either-$R$-or-$S$" relation with confidence which is equal to the larger of $R(t)^+$ and $S(t)^+$. Using the same logic, we can now believe in the absence of the tuple $t$ from the "either-$R$-or-$S$" relation with confidence which is equal to the smaller (because $t$ must be absent from both $R$ and $S$ for it to be absent from the union) of $R(t)^-$ and $S(t)^-$. The definition of *complement* and of all the other operators on neutrosophic relations defined later can (and should) be understood in the same way.

**Proposition 3** The operators $\hat{\cup}$ and unary $\hat{-}$ on neutrosophic relations are strong generalizations of the operators $\cup$ and unary $-$ on fuzzy relations.

**Proof** Let $R$ and $S$ be consistent neutrosophic relations on scheme $\Sigma$. Then $reps_\Sigma(R \hat{\cup} S)$ is the set

$$\{Q \mid \wedge_{t_i \in \tau(\Sigma)} (\max\{R(t_i)^+, S(t_i)^+\} \leq Q(t_i) \leq 1 - \min\{R(t_i)^-, S(t_i)^-\})\}$$

This set is the same as the set

$$\{r \cup s \mid \wedge_{t_i \in \tau(\Sigma)} (R(t_i)^+ \leq r(t_i) \leq 1 - R(t_i)^-), \wedge_{t_i \in \tau(\Sigma)} (S(t_i)^+ \leq s(t_i) \leq 1 - S(t_i)^-)\}$$

which is $S(\cup)(reps_\Sigma(R), reps_\Sigma(S))$. Such a result for unary $\hat{-}$ can also be shown similarly.

For sake of completeness, we define the following two related set-theoretic operators:

**Definition 19** Let $R$ and $S$ be neutrosophic relations on scheme $\Sigma$. Then,

(a) the *intersection* of $R$ and $S$, denoted $R\hat{\cap}S$, is a neutrosophic relation on scheme $\Sigma$, given by

$$(R\hat{\cap}S)(t) = \langle \min\{R(t)^+, S(t)^+\}, \max\{R(t)^-, S(t)^-\} \rangle, \text{ for any } t \in \tau(\Sigma).$$

(b) the *difference* of $R$ and $S$, denoted $R\hat{-}S$, is a neutrosophic relation on scheme $\Sigma$, given by

$$(R\hat{-}S)(t) = \langle \min\{R(t)^+, S(t)^-\}, \max\{R(t)^-, S(t)^+\} \rangle, \text{ for any } t \in \tau(\Sigma).$$

The following proposition relates the intersection and difference operators in terms of the more fundamental set-theoretic operators union and complement.

**Proposition 4** For any neutrosophic relations $R$ and $S$ on the same scheme, we have

$$R\hat{\cap}S = \hat{-}(\hat{-}R\hat{\cup}\hat{-}S), \text{ and}$$

$$R\hat{-}S = \hat{-}(\hat{-}R\cup S).$$

Proof By definition,

$$\hat{-}R(t) = \langle R(t)^-, R(t)^+ \rangle$$

$$\hat{-}S(t) = \langle S(t)^-, S(t)^+ \rangle$$

and $(\hat{-}R\hat{\cup}\hat{-}S)(t) = \langle \max(R(t)^-, S(t)^-), \min(R(t)^+, S(t)^+) \rangle$

so, $(\hat{-}(\hat{-}R\hat{\cup}\hat{-}S))(t) = \langle \min(R(t)^+, S(t)^+), \max(R(t)^-, S(t)^-) \rangle = R\hat{\cap}S(t).$

The second part of the result can be shown similarly.

## 4.2 Relation-Theoretic Operators

We now define some relation-theoretic algebraic operators on neutrosophic relations.

**Definition 20**   Let $R$ and $S$ be neutrosophic relations on schemes $\Sigma$ and $\Delta$, respectively. Then, the *natural join* (further for short called *join*) of $R$ and $S$, denoted $R \stackrel{\wedge}{\bowtie} S$, is a neutrosophic relation on scheme $\Sigma \cup \Delta$, given by

$$(R \stackrel{\wedge}{\bowtie} S)(t) = \langle \min\{R(\pi_\Sigma(t))^+, S(\pi_\Delta(t))^+\}, \max\{R(\pi_\Sigma(t))^-, S(\pi_\Delta(t))^-\} \rangle,$$

where $\pi$ is the usual projection of a tuple.

It is instructive to observe that, similar to the intersection operator, the minimum of the belief factors and the maximum of the doubt factors are used in the definition of the join operation.

**Proposition 5**   $\stackrel{\wedge}{\bowtie}$ is a strong generalization of $\bowtie$.

Proof   Let $R$ and $S$ be consistent neutrosophic relations on schemes $\Sigma$ and $\Delta$, respectively. Then $reps_{\Sigma \cup \Delta}(R \stackrel{\wedge}{\bowtie} S)$ is the set $\{Q \in F(\Sigma \cup \Delta) \mid \wedge_{t_i \in \tau(\Sigma \cup \Delta)} (\min\{R_{\pi_\Sigma}(t_i)^+, S_{\pi_\Delta}(t_i)^+\} \leq Q(t_i) \leq 1 - \max\{R_{\pi_\Sigma}(t_i)^-, S_{\pi_\Delta}(t_i)^-\})\}$ and $S(\bowtie)(reps_\Sigma(R), reps_\Delta(S)) = \{r \bowtie S \mid r \in reps_\Sigma(R), s \in reps_\Delta(S)\}$.

Let $Q \in reps_{\Sigma \cup \Delta}(R \stackrel{\wedge}{\bowtie} S)$. Then $\pi_\Sigma(Q) \in reps_\Sigma(R)$, where $\pi_\Sigma$ is the usual projection over $\Sigma$ of fuzzy relations. Similarly, $\pi_\Delta(Q) \in reps_\Delta(R)$, Therefore, $Q \in S(\bowtie)(reps_\Sigma(R), reps_\Delta(S))$.

Let $Q \in S(\bowtie)(reps_\Sigma(R), reps_\Delta(S))$. Then $Q(t_i) \geq \min\{R_{\pi_\Sigma}(t_i)^+, S_{\pi_\Delta}(t_i)^+\}$ and

$Q(t_i) \leq \min\{1 - R_{\pi_\Sigma}(t_i)^-, 1 - S\pi_\Delta(t_i)^-\} = 1 - \max\{R_{\pi_\Sigma}(t_i)^-, S_{\pi_\Delta}(t_i)^-\}$,   for any   $t_i \in \tau(\Sigma \cup \Delta)$,

because $R$ and $S$ are consistent. Therefore, $Q \in reps_{\Sigma \cup \Delta}(R \stackrel{\wedge}{\bowtie} S)$.

We now present the projection operator.

**Definition 21**   Let $R$ be a neutrosophic relation on scheme $\Sigma$, and $\Delta \subseteq \Sigma$. Then, the *projection* of $R$ onto $\Delta$, denoted $\stackrel{\wedge}{\pi}_\Delta(R)$, is a neutrosophic relation on scheme $\Delta$, given by

$$(\hat{\pi}_\Delta(R))(t) = \langle \max\{R(u)^+ \mid u \in t^\Sigma\}, \min\{R(u)^- \mid u \in t^\Sigma\}\rangle.$$

The belief factor of a tuple in the projection is the maximum of the belief factors of all of the tuple's extensions onto the scheme of the input neutrosophic relation. Moreover, the doubt factor of a tuple in the projection is the minimum of the doubt factors of all of the tuple's extensions onto the scheme of the input neutrosophic relation.

We present the selection operator next.

**Definition 22** Let $R$ be a neutrosophic relation on scheme $\Sigma$, and let $F$ be any logic formula involving attribute names in $\Sigma$, constant symbols (denoting values in the attribute domains), equality symbol $=$, negation symbol $\neg$, and connectives $\vee$ and $\wedge$. Then, the *selection* of $R$ by $F$, denoted $\hat{\sigma}_F(R)$, is a neutrosophic relation on scheme $\Sigma$, given by

$$(\hat{\sigma}_F(R))(t) = \langle \alpha, \beta \rangle, \text{ where}$$

$$\alpha = \begin{cases} R(t)^+ & \text{if } t \in \sigma_F(\tau(\Sigma)) \\ 0 & \text{otherwise} \end{cases} \text{ and } \beta = \begin{cases} R(t)^- & \text{if } t \in \sigma_F(\tau(\Sigma)) \\ 1 & \text{otherwise} \end{cases}$$

where $\sigma_F$ is the usual selection of tuples satisfying $F$ from ordinary relations.

If a tuple satisfies the selection criterion, its belief and doubt factors are the same in the selection as in the input neutrosophic relation. In the case where the tuple does not satisfy the selection criterion, its belief factor is set to 0 and the doubt factor is set to 1 in the selection.

**Proposition 6** The operators $\hat{\pi}$ and $\hat{\sigma}$ are strong generalizations of $\pi$ and $\sigma$, respectively.

Proof Similar to that of Proposition 5.

**Example 2** Relation schemes are sets of attribute names, but in this example we treat them as ordered sequence of attribute names (which can be obtained through permutation of attribute names), so tuples can be viewed as the usual lists of values. Let $\{a,b,c\}$ be a common domain for all attribute names, and let $R$ and $S$ be the following neutrosophic relations on schemes $\langle X,Y \rangle$ and $\langle Y,Z \rangle$ respectively.

| t | R(t) |
|---|---|
| (a,a) | <0,1> |
| (a,b) | <0,1> |
| (a,c) | <0,1> |
| (b,b) | <1,0> |
| (b,c) | <1,0> |
| (c,b) | <1,1> |

| t | S(t) |
|---|---|
| (a,c) | <1,0> |
| (b,a) | <1,1> |
| (c,b) | <0,1> |

For other tuples which are not in the neutrosophic relations $R(t)$ and $S(t)$, their $\langle \alpha, \beta \rangle = \langle 0,0 \rangle$ which means no any information available. Because $R$ and $S$ are inconsistent, we first use split operation to transform them into pseudo-consistent and apply the relation-theoretic operations on them and transform the result back to arbitrary neutrosophic set using combine operation. Then, $T_1 = \nabla(\Delta(R) \stackrel{\wedge}{\bowtie} \Delta(S))$ is a neutrosophic relation on scheme $\langle X, Y, Z \rangle$ and $T_2 = \nabla(\hat{\pi}_{\langle X,Z \rangle}(\Delta(T_1)))$ and $T_3 = \hat{\sigma}_{X \neg = Z}(T_2)$ are neutrosophic relations on scheme $\langle X, Z \rangle$. $T_1$, $T_2$, and $T_3$ are shown below:

| t | $T_1(t)$ |
|---|---|
| (a,a,a) | <0,1> |
| (a,a,b) | <0,1> |
| (a,a,c) | <0,1> |

| (a,b,a) | <0,1> |
| (a,b,b) | <0,1> |
| (a,b,c) | <0,1> |
| (a,c,a) | <0,1> |
| (a,c,b) | <0,1> |
| (a,c,c) | <0,1> |
| (b,b,a) | <1,1> |
| (b,c,b) | <0,1> |
| (c,b,a) | <1,1> |
| (c,b,b) | <0,1> |
| (c,b,c) | <0,1> |
| (c,c,b) | (<0,1> |

| $t$ | $T_2(t)$ |
|---|---|
| (a,a) | <0,1> |
| (a,b) | <0,1> |
| (a,c) | <0,1> |
| (b,a) | <1,0> |
| (c,a) | <1,0> |

| $t$ | $T_3(t)$ |
|---|---|
| (a,a) | <0,1> |
| (a,b) | <0,1> |
| (a,c) | <0,1> |
| (b,a) | <1,0> |
| (b,b) | <0,1> |
| (c,a) | <1,0> |
| (c,c) | <0,1> |

## 5. An Application

Consider the target recognition example presented in [36]. Here, an autonomous vehicle needs to identify objects in a hostile environment such as a military battlefield. The autonomous vehicle is equipped with a number of sensors which are used to collect data, such as speed and size of the objects (tanks) in the battlefield. Associated with each sensor, we have a set of rules that describe the type of the object based on the properties detected by the sensor.

Let us assume that the autonomous vehicle is equipped with three sensors resulting in data collected about radar readings, of the tanks, their gun characteristics, and their speeds. What follows is a set of rules that associate the type of object with various observations.

**Radar Readings**:

- Reading $r_1$ indicates that the object is a T-72 tank with belief factor 0.80 and doubt factor 0.15.

- Reading $r_2$ indicates that the object is a T-60 tank with belief factor 0.70 and doubt factor 0.20.

- Reading $r_3$ indicates that the object is not a T-72 tank with belief factor 0.95 and doubt factor 0.05.

- Reading $r_4$ indicates that the object is a T-80 tank with belief factor 0.85 and doubt factor 0.10.

**Gun Characteristics**:

- Characteristic $c_1$ indicates that the object is a T-60 tank with belief factor 0.80 and doubt factor 0.20.

- Characteristic $c_2$ indicates that the object is not a T-80 tank with belief factor 0.90 and doubt factor 0.05.

- Characteristic $c_3$ indicates that the object is a T-72 tank with belief factor 0.85 and doubt factor 0.10.

**Speed Characteristics**:

- Low speed indicates that the object is a T-60 tank with belief factor 0.80 and doubt factor 0.15.

- High speed indicates that the object is not a T-72 tank with belief factor 0.85 and doubt factor 0.15.

- High speed indicates that the object is not a T-80 tank with belief factor 0.95 and doubt factor 0.05.

- Medium speed indicates that the object is not a T-80 tank with belief factor 0.80 and doubt factor 0.10.

These rules can be captured in the following three neutrosophic relations:

Radar Rules

| Reading | Object | Confidence Factors |
|---|---|---|
| $r_1$ | T-72 | <0.80,0.15> |
| $r_2$ | T-60 | <0.70,0.20> |
| $r_3$ | T-72 | <0.05,0.95> |
| $r_4$ | T-80 | <0.85,0.10> |

Gun Rules

| Reading | Object | Confidence Factors |
|---|---|---|
| $c_1$ | T-60 | <0.80,0.20> |
| $c_2$ | T-80 | <0.05,0.90> |
| $c_3$ | T-72 | <0.85,0.10> |

Speed Rules

| Reading | Object | Confidence Factors |
|---|---|---|
| low | T-60 | <0.80,0.15> |
| high | T-72 | <0.15,0.85> |
| high | T-80 | <0.05,0.95> |
| medium | T-80 | <0.10,0.80> |

The autonomous vehicle uses the sensors to make observations about the different objects and then uses the rules to determine the type of each object in the battlefield. It is quite possible that two different sensors may identify the same object as of different types, thereby introducing inconsistencies.

Let us now consider three objects $o_1$, $o_2$ and $o_3$ which need to be identified by the autonomous vehicle. Let us assume the following observations made by the three sensors about the three objects. Once again, we assume certainty factors (maybe derived from the accuracy of the sensors) are associated with each observation.

Radar Data

| Object-id | Reading | Confidence Factors |
|---|---|---|
| $o_1$ | $r_3$ | <1.00,0.00> |
| $o_2$ | $r_1$ | <1.00,0.00> |
| $o_3$ | $r_4$ | <1.00,0.00> |

Gun Data

| Object-id | Reading | Confidence Factors |
|---|---|---|
| $o_1$ | $c_3$ | <0.80,0.10> |
| $o_2$ | $c_1$ | <0.90,0.10> |
| $o_3$ | $c_2$ | <0.90,0.10> |

Speed Data

| Object-id | Reading | Confidence Factors |
|---|---|---|
| $o_1$ | high | <0.90,0.10> |
| $o_2$ | low | <0.95,0.05> |
| $o_3$ | medium | <0.80,0.20> |

Given these observations and the rules, we can use the following algebraic expression to identify the three objects:

$$\hat{\pi}_{Object-id, Ojbect}(Radar\ Data\ \hat{\infty}\ Radar\ Rules)\hat{\cap}$$
$$\hat{\pi}_{Object-id, Object}(Gun\ Data\ \hat{\infty}\ Gun\ Rules)\hat{\cap}$$
$$\hat{\pi}_{Object-id, Object}(Speed\ Data\ \hat{\infty}\ Speed\ Rules)$$

The intuition behind the intersection is that we would like to capture the common (intersecting) information among the three sensor data. Evaluating this expression, we get the following neutrosophic relation:

| Object-id | Reading | Confidence Factors |
|---|---|---|
| $o_1$ | T-72 | <0.05,0.00> |
| $o_2$ | T-80 | <0.00,0.05> |
| $o_3$ | T-80 | <0.05,0.00> |

It is clear from the result that by the given information, we could not infer any useful information that is we could not decide the status of objects $o_1$, $o_2$ and $o_3$.

## 6. Conclusions and Future Work

We have presented a generalization of fuzzy relations, intuitionistic fuzzy relations (interval-valued fuzzy relations), and paraconsistent relations, called neutrosophic relations, in which we allow the representation of confidence (belief and doubt) factors with each tuple. The algebra on fuzzy relations is appropriately generalized to manipulate neutrosophic relations.

Various possibilities exist for further study in this area. Recently, there has been some work in extending logic programs to involve quantitative paraconsistency. Paraconsistent logic programs were introduced in [37] and probabilistic logic programs in [38]. Paraconsistent logic programs allow negative atoms to appear in the head of clauses (thereby resulting in the possibility of dealing with inconsistency), and probabilistic logic programs associate confidence measures with literals and with entire clauses. The semantics of these extensions of logic programs have already been presented, but implementation strategies to answer queries have not been discussed. We propose to use the model introduced in this paper in computing the semantics of these extensions of logic programs. Exploring application areas is another important thrust of our research.

We developed two notions of generalizing operators on fuzzy relations for neutrosophic relations. Of these, the stronger notion guarantees that any generalized operator is "well-behaved" for neutrosophic relation operands that contain consistent information.

For some well-known operators on fuzzy relations, such as union, join, and projection, we introduced generalized operators on neutrosophic relations. These generalized operators maintain the belief system intuition behind neutrosophic relations, and are shown to be "well-behaved" in the sense mentioned above.

Our data model can be used to represent relational information that may be incomplete and inconsistent. As usual, the algebraic operators can be used to construct queries to any database systems for retrieving vague information.